\documentclass[prd,aps,twocolumn,tightenlines,superscriptaddress,nofootinbib,preprintnumbers,english]{revtex4}
\usepackage[T1]{fontenc}
\usepackage[latin9]{inputenc}
\usepackage{amsmath}
\usepackage{amssymb}
\usepackage{graphicx}
\usepackage{esint}
\usepackage{babel}
\usepackage{color}
\usepackage[usenames,dvipsnames]{xcolor}
\usepackage{tcolorbox}
\usepackage[colorlinks,linkcolor=blue,anchorcolor=blue,citecolor=blue]{hyperref}

\newcommand{\JW}[1]{[\textcolor{blue}{JW: #1}]}
\newcommand{\Red}[1]{\textcolor{red}{#1}}
\newcommand{\Blue}[1]{\textcolor{blue}{#1}}

\usepackage[capitalize]{cleveref}
\usepackage{subfig}

\begin{document}

\title{On the Renormalization of  Entanglement Entropy}

\author{Jiunn-Wei Chen}

\email{jwc@phys.ntu.edu.tw}
\affiliation{Department of Physics, Center for Theoretical Sciences, and Leung Center for Cosmology and Particle Astrophysics, National Taiwan University, Taipei, Taiwan 106}
\affiliation{Helmholtz-Institut f\"{u}r Strahlen- und Kernphysik and Bethe Center
for Theoretical Physics, Universit\"{a}t Bonn, D-53115 Bonn, Germany}

\author{Jin-Yi Pang}
\email{pang@hiskp.uni-bonn.de}
\affiliation{Helmholtz-Institut f\"{u}r Strahlen- und Kernphysik and Bethe Center
for Theoretical Physics, Universit\"{a}t Bonn, D-53115 Bonn, Germany}

\begin{abstract}

The renormalization of entanglement entropy of quantum field theories is investigated in the simplest setting with a $\lambda \phi^4$ scalar field theory. The 3+1 dimensional spacetime is separated into two regions by an infinitely flat 2-dimensional interface. The entanglement entropy of the system across the interface has an elegant geometrical interpretation using the replica trick, which requires putting the field theory on a curved spacetime background. We demonstrate that the theory, and hence the entanglement entropy, is renormalizable at order $\lambda$ once all the relevant operators up to dimension-4 are included in the action. This exercise has a one-to-one correspondence to entanglement entropy interpretation of the black hole entropy which suggests that our treatment is sensible. Our study suggests that entanglement entropy is renormalizable and is a physical quantity.

\end{abstract}
\maketitle

\section{Introduction}

An interesting attempt to understand the Bekenstein-Hawking formula of black hole entropy $S_{BH}=A/4G$, with $A$ the black hole horizon area and $G$ the gravitational constant, is to relate it to the entanglement entropy ($S_E$)  across the black hole horizon \cite{tHooft:1984kcu,Bombelli:1986rw,Srednicki:1993im,Frolov:1993ym,Das:2008sy}. $S_E$ quantifies the entanglement between the degrees of freedom inside and outside the black hole. But first attempts to compute $S_E$ from free fields yielded 
the desired $A$ dependence whereas the prefactor is divergent \cite{Bombelli:1986rw,Srednicki:1993im,Frolov:1993ym}. Susskind and Uglum suggested that the divergence just renormalizes the bare gravitational constant $G$ to the renormalized one, $G_R$ such that  $S_{BH}=S_E=A/4G_R$ \cite{Susskind:1993ws,Susskind:1994sm}. This suggestion was confirmed by explicit computations in the massive black hole limit for free fields while treating gravity classical. Whether this result is modified by the finite black hole mass or terms beyond Einstein gravity or when gravity is quantized are interesting questions for further exploration  \cite{Callan:1994py,Kabat:1994vj,Solodukhin:1994yz,Fursaev:1994pq,Demers:1995dq,Kabat:1995eq,Larsen:1995ax,Fursaev:1996uz,Cooperman:2013iqr}.

In a condense matter system, $S_E$ can still be defined as the entanglement of degrees of freedom across the interface of area $A$ between two regions\cite{Vidal2003,Latorre2004}. 
But then how important is gravity to the determination of $S_E$? 
Is $S_E$ still renormalizable without gravity in the theory? If not, why is gravity so special and why is it so important even for a condensed matter system? If yes, how is $S_E$ renormalized?
A popular treatment in the computation of $S_E$ is just imposing a UV cut-off  
without renormalizing it. In this treatment, $S_E/A$ is set by the cut-off scale and when the UV cut-off increases, $S_E$ also increases since there are more degrees of freedom that can entangle. However, this does not always lead to a positive $S_E$ \cite{Kabat:1995eq,Larsen:1995ax,Fursaev:1996uz,Chen:2015kfa}, which is required by definition. Different ways to fix the problem usually lead to different results. 
Therefore, it is important to ask a more fundamental question, is $S_E$ a physical quantity? If yes, how do we renormalize it? 

In this work, we try to address this issue by renormalizing $S_E$ in the simplest example.  We set up the problem using the ``replica trick'' which provides a geometrical interpretation of $S_E$ \cite{Callan:1994py} which we will review in the following section. Then for a 3+1 dimensional scalar field theory with quartic interaction (a $\lambda \phi^4$ theory), we compute the entanglement entropy for two regions separated by a flat infinite two dimensional plan to order $\lambda$. The corresponding geometry of the Euclidean spacetime in replica trick is $C_{\epsilon} \times R^2$, where $C_{\epsilon}$ is a two dimensional cone with a deficit angle $\epsilon$. And then $S_E$ is the linear response of the effective action to a vanishing deficit angle. Hence the problem of $S_E$ computation is nothing but a field theory problem in a curve spacetime. This spacetime does not have the Poincare symmetry (or translational and rotational symmetries in Euclidean space) of the flat space time. Hence, to renormalize the theory, we write down all the renormalizable terms
with mass dimensions $\le 4$ and with symmetries satisfied by the $C_{\epsilon} \times R^2$ space. We derived the required scalar propagator in this theory and show that all divergent diagrams at $\mathcal{O}(\epsilon \lambda)$ can all be renormalized and so is $S_E$ at this order. Finally, we do the same exercise to the black hole case by replacing the curve spacetime back ground by gravity and find that there is a one-to-one correspondence between the non-gravitational theory and the black hole case mathematically. While this result suggests that our formulation is sensible from the point of view of general relativity, the deficit angle in a condensed matter system in flat space remains an illusive concept worth further exploration.


\section{Review of the Replica Trick}

Suppose our system occupies an infinitely large and flat $3+1=4$ dimensional spacetime, with
the three dimensional space divided into
two regions $V$ and $\bar{V}$ by a time independent, flat 2-dimensional plan.
Then the entanglement entropy, $S_E$, of
a quantum theory between the two subregions is defined by the von
Neumann entropy of the reduced density matrix $\rho_V=\text{tr}_{\bar{V}}[ \rho]$ by tracing out the degrees of freedom in region $\bar{V}$:
\begin{equation}
\label{S_E}
S_{E}=-\text{tr}[\rho_{V}\ln\rho_{V}]=-\left.\frac{\partial}{\partial n}\ln \text{tr}[\rho_{V}^{n}]\right\vert _{n\to1} ,
\end{equation}
where $\text{tr}[\rho_{V}]$=1 is used.
This expression
is called the replica trick because it involves $n$ copies of $\rho_{V}$.

\begin{figure}
\begin{centering}
\subfloat[]{\includegraphics[scale=0.6]{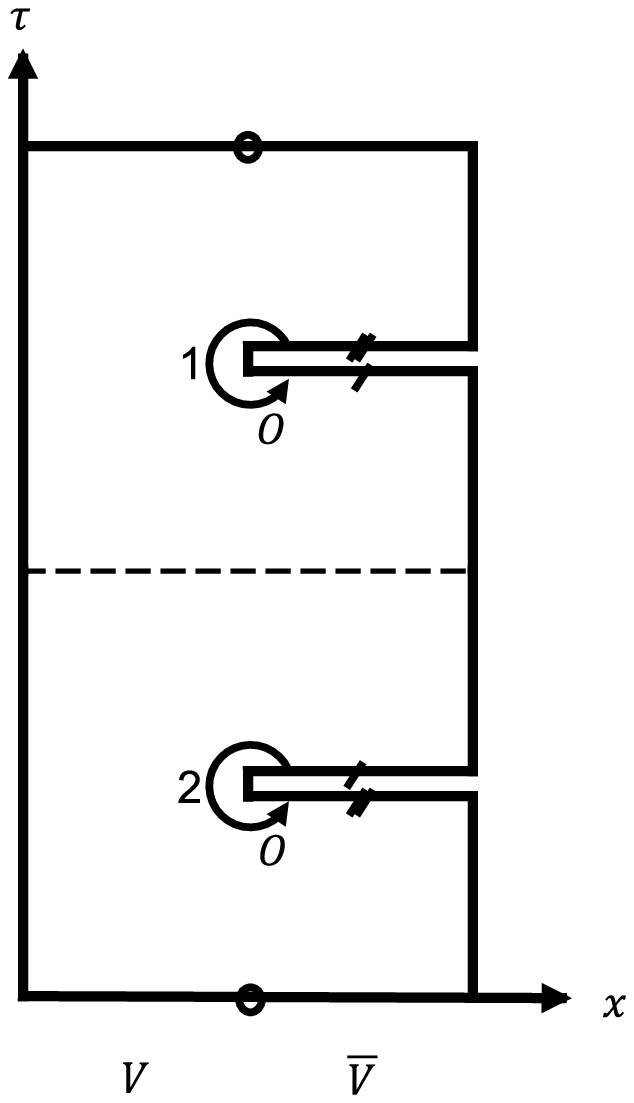}\label{fig:replica_b}

}\quad\quad\subfloat[]{\includegraphics[scale=0.6]{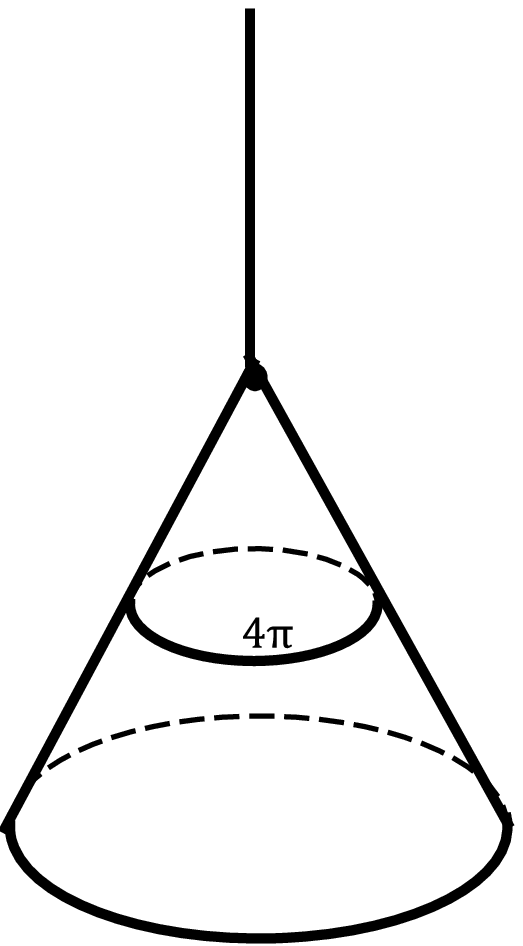}\label{fig:replica_c}

}
\par\end{centering}
\captionsetup{justification=raggedright,singlelinecheck=false}
\caption
{$\text{tr}[\rho_{V}^{2}]$ can be computed by the 2-sheeted 2-sheeted Riemann surface shown in (a). The period of $\tau$ in region $V$ is twice of that in region $\bar{V}$ (see the text). A close loop around point $O$ goes through contours 1 and 2 with a total angle $4 \pi$. For a system of infinite spacetime volume, (a) can be drawn as (b) with a 2-dimensional cone of angle $4 \pi$ and 2 codimensions. The line connecting to the tip of the cone in (b) denotes the 2-dimensional brane which is the interface. The replica trick requires generalizing this picture to $\text{tr}[\rho_{V}^{n}]$ with $n \to 1$.}

\end{figure}

An elegant path integral formulation to compute the entanglement entropy
using the replica trick was introduced in \cite{Callan:1994py}.
In this set up, one recalls that
$\rho_{ij}\propto \left\langle i\left\vert e^{-H/T} \right\vert j\right\rangle $ for a thermal equilibrium system with Hamiltonian $H$ and temperature $T$. 
$Z=\text{tr}[\rho]$ is the partition function calculated in finite temperature
field theory with the range of  Euclidean time $\tau=[0,1/T]$ and with 
appropriate boundary conditions: $\phi(\tau=0,\mathbf{x})=\pm \phi(\tau=1/T,\mathbf{x})$ with the $+(-)$ sign if $\phi$ is a boson(fermion) field.
Then tr$[\rho^{2}]\propto$ tr$[e^{-2H/T}]$ can be computed by doubling the period in $\tau$ such that boundary conditions are imposed at $\tau=0$ and $2/T$. Similarly,
tr$[\rho_{V}^{2}]$ is computed by doubling the period in $\tau$ for region $V$ while keeping a single
period in region $\bar{V}$. This can be shown as a 2-sheeted Riemann surface
as in Fig.(\ref{fig:replica_b}), where we have the period $\tau = [0,2/T]$ in region $V$ and periods $\tau = [1/2T,3/2T]$ and $\tau = [3/2T,5/2T \sim -1/2T]$ in region $\bar{V}$. In this figure, only $\tau$ (the vertical direction) and the direction perpendicular to the interface (the horizontal direction) are shown while the 2-dimensional interface in the perpendicular direction is not shown.
If we circle around point $O$ by contour 1, then it will connect to contour 2 with the total angle for a closed loop to be $4 \pi$. 

This analysis can be generalized to tr$[\rho_{V}^{n}]$ for an arbitrary integer $n$ for arbitrary sizes of $V$, $\bar{V}$, and $1/T$.  As a result, $\text{tr}[\rho_{V}^{n}]$ becomes the partition function
$Z_{n}$ on the $n$-sheeted Riemann surface
normalized by $Z_{1}^{n}$ which follows from imposing the normalization $\text{tr}[\rho_{V}]$=1: 
\begin{equation}
\text{tr}[\rho_{A}^{n}]=\frac{Z_{n}}{Z_{1}^{n}} .
\label{eq:replica1}
\end{equation}
Then Eq.(\ref{S_E}) yields
\begin{equation}
\label{SE}
S_{E}=\left.\left(-\frac{\partial}{\partial n}+1\right)\ln Z_{n}\right|_{n\to1} .
\end{equation}

In this work, we concentrate on the simplest case with the sizes of $V$, $\bar{V}$, and $1/T$ all become infinite ($T=0$) and the interface between $V$ and $\bar{V}$ is a flat infinite plane. In this limit, the $n$-sheeted Riemann surface in Fig.(\ref{fig:replica_b}) can be redrawn to Fig.(\ref{fig:replica_c}) which has the $C_{\epsilon} \times R^2$ topology. The Euclidean time and the the direction perpendicular to the interface form the 2-dimensional cone $C_{\epsilon}$ with the deficit angle $\epsilon$ of the cone satisfing $n=1-\epsilon/2\pi$. $R^2$ is the space parallel to the 2-dimensional interface but transverse to the cone. The line connecting to the tip of the cone in Fig.(\ref{fig:replica_c}) denotes the 2-dimensional brane which is the interface.

%
%
%
%
Now Eq.(\ref{SE}) becomes
\begin{equation}
\label{SE2}
S_{E}=\left.\left(2 \pi \frac{\partial}{\partial \epsilon}+1\right)\ln Z(\epsilon)\right|_{\epsilon \to 0} ,
\end{equation}
with $S_E$ probing the linear response of the partition function to the deficit angel.

Thermal entropy is a special case of entanglement entropy with vanishing $\bar{V}$.  
As shown in Fig.(\ref{fig:replica_b}), without $\bar{V}$, the spacetime topology becomes $S^1 \times R^3$ in flat Euclidean space. $n$ controls the size of $S^1$ which is $\beta=n/T$. Then Eq.(\ref{SE}) yields the standard equation for thermal entropy \begin{equation}
\label{S_th}
S=\left(-\beta \frac{\partial}{\partial \beta}+1\right)\ln Z .
\end{equation}

The situation changes when it comes to the thermal entropy of a black hole. The Euclidean spacetime geometry just outside the blackhole horizon is $R^2 \times S^2$. The period of the polar angle in $R^2$ is proportional to $\beta$. When $\beta$ is not equal to the inverse Hawking temperature of the black hole $\beta_H$, $R^2$ becomes a cone of deficit angel $\epsilon=2 \pi (\beta_H-\beta)/\beta_H$. Then Eq.(\ref{S_th}) becomes  
\begin{equation}
\label{S_BH}
S_{BH}=\left.\left(2 \pi \frac{\partial}{\partial \epsilon}+1\right)\ln Z(\epsilon)\right|_{\epsilon \to 0} ,
\end{equation}
which is very similar to Eq.(\ref{SE2}).  

\section{Renormalized Entanglement Entropy in the $\lambda \phi^4$ Theory}

To compute the entanglement entropy $S_E$ using Eq.(\ref{SE}), we need to construct the theory in the $C_{\epsilon} \times R^2$ space shown in Fig.(\ref{fig:replica_c}). 
The interface is shown as a two dimensional brane ($R^2$) which has a co-dimension two cone $C_{\epsilon}$ outside the brane. The space is locally flat outside of the brane. There could be fields only live on the brane. 

A $\lambda \phi^4$ theory in this space has the Lagrangian
\begin{eqnarray}
\mathcal{L} & = & \frac{1}{2}(\partial\phi)^{2}+\frac{1}{2}m^{2}\phi^{2}+\frac{\lambda}{4}\phi^{4} \nonumber\\
& + & \frac{1}{2}Z_{\phi}(\partial\phi)^{2}+\frac{1}{2}Z_{m}\phi^{2}+\frac{1}{4}Z_{\lambda}\phi^{4}+Z_4  \nonumber\\
 & + & \epsilon\delta^{(2)}(x_{\parallel})\left(Z_2+Z_0 \phi^{2}\right)+\mathcal{O}(\epsilon^{2}) .
\label{eq:o(n)_sigma_Lagrangian}
\end{eqnarray}
The first two lines are the usual Lagrangian in flat space with the full Poincare symmetry. $m$ and $\lambda$ are renormalized quantities. Terms in the second line are conterterms.  $Z_4$ is the counterterm that renormalizes the cosmological constant. Terms in the third line breaks the Poincare symmetry due to the 2d brane. The $Z_2$ term is the 2d brane tension while the $Z_0$ 
is the brane coupling to the scalar. The third line should vanish when $\epsilon \to 0$. Hence under Taylor expansion, their couplings are proportional to $\epsilon$ for small $\epsilon$. (The first derivative of $\epsilon$ has been assumed to exist when we apply the replica trick.)  This system is nothing but a quantum field theory in a curved spacetime background. We expect that once all the relevant operators (i.e. operators up to dimension four for a weakly coupled system) with the symmetries of the problem are included, then the theory should be renormalizable. While the renormalizability is known to all orders in $\lambda$ at $\mathcal{O}(\epsilon^0)$, 
we will demonstrate that this is also the case at $\mathcal{O}(\epsilon \lambda)$.

The Green's function $G_{n}(x,x')$ for the free scalar field on a $n$-sheeted Riemann surface satisfies
\begin{eqnarray}
 (-\partial^2+m^2)G_{n}(x,x')=\delta^4(x-x') .
\end{eqnarray}
The conical singularity breaks translational symmetry such that $G_{n}(x,x')$ depends on both $x$ and $x'$ instead of $x-x'$ alone. As shown in Refs. \cite{Calabrese2004,Hertzberg2013} and in the Appendix, the Green function has the solution
\begin{eqnarray}
\label{Gn}
 G_{n}(x,x') & = & \int\frac{d^{2}p_{\perp}}{(2\pi)^{2}}\int qdq\frac{e^{ip_{\perp}(x_{\perp}-x_{\perp}')}}{q^{2}+p_{\perp}^{2}+m^{2}} \left[J_{0}(qr)J_{0}(qr')\right. \nonumber \\
 &  &\left.+2\sum_{l=1}^{\infty}J_{\frac{l}{n}}(qr)J_{\frac{l}{n}}(qr')\cos\frac{l}{n}(\theta-\theta')\right]
\end{eqnarray}
where polar coordinates $(r,\theta)$ and $(r',\theta')$ are used to describe points
on longitudinal plane with respect to the conical singularity. 

Under the $\epsilon$ expansion, we have
\begin{eqnarray}
G_{1+\epsilon}(x,x') & = & G(x-x')+\epsilon f(x,x')+\mathcal{O}(\epsilon^2),
\end{eqnarray}
where we have rewritten the Green's function in flat space $G_1(x,x')$ as $G(x-x')$ since translational symmetry is satisfied for $G_1(x,x')$. 

There are two classes of diagrams at $\mathcal{O}(\epsilon)$. The first one is with one $\mathcal{O}(\epsilon)$ coupling insertion and with all the other couplings and propagators of $\mathcal{O}(\epsilon^0)$. The second one is with one $ f(x,x')$ propagator but with all the other propagators and couplings of $\mathcal{O}(\epsilon^0)$, which implies if the $f(x,x')$ propagator is removed, then the rest of the diagram is translational invariant. Therefore, a generic diagram of the second class can be expressed as
\begin{eqnarray}
\label{11}
\int dx dx' f(x,x')F(x-x') .
\end{eqnarray}
With this condition, we show in the Appendix that the $\mathcal{O}(\epsilon)$ propagator in the Fourier space can be effectively written as:
\begin{eqnarray}
\label{fn}
f(p) \to \frac{A_{\perp}}{12} \frac{\delta^2(p_{\parallel})}{p_{\perp}^2+m^2} ,
\end{eqnarray}
where $A_{\perp}$ is the area of the interface.
This expression has a mass dimension minus six because $G_{n}(x,x')$ has a mass dimension two. 
Now we see all the couplings and propagators at $\mathcal{O}(\epsilon)$ are proportional to $A_{\perp}$, so every diagram at $\mathcal{O}(\epsilon)$ is  proportional to $A_{\perp}$ as well.


The $\mathcal{O}(\epsilon)$ propagator of Eq.(\ref{fn}) has two powers of momentum lass than the $\mathcal{O}(\epsilon^0)$ propagator. Hence, the standard power counting analysis shows that at $\mathcal{O}(\epsilon)$, only the two point functions and zero point functions are divergent and need to be renormalized. The two point functions could diverge logarithmically while the zero point function could diverge quadratically at $\mathcal{O}(\epsilon)$. 

\begin{figure}
\begin{centering}
\includegraphics[scale=0.3]{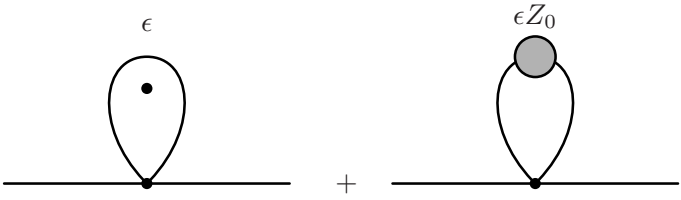}
\par\end{centering}
\captionsetup{justification=raggedright,singlelinecheck=false}
\caption
{Two point functions at $O(\epsilon\lambda)$. The left diagram has an $O(\epsilon)$ propagator circling the conical singularity (or the 2d brane) denoted by the dot. The right diagram has one insertion of the $Z_{0}$ coupling. }
\label{fig:Two-point-function}

\end{figure}

We will start from the renormalization of two point functions at $\mathcal{O}(\epsilon \lambda)$ shown in Fig.(\ref{fig:Two-point-function}). The sum of the two diagrams is proportional to 
\begin{eqnarray}
\label{2pt}
-\frac{1}{12}\int\frac{d^{2}p}{(2\pi)^{2}}\frac{1}{p^{2}+m^{2}}
 +  Z_0\int\frac{d^{4}p}{(2\pi)^{4}}\frac{1}{(p^{2}+m^{2})^{2}} .
\end{eqnarray}
Then using $d^d p = dp p^{d-1}\Omega_d$, $\Omega_4=2 \pi^2$ and $\Omega_2=2 \pi$, the two point function can be renormalized by setting $Z_0=\pi/3$ such that the divergence between the two diagrams cancel and (\ref{2pt}) equals $1/(48 \pi)$. Here we do not demand the first diagram being renormalized by the tree diagram with one insertion of  $Z_0$ because in general $Z_0$ is of order $\lambda^0$, not  $\lambda$.

\begin{figure}
\begin{centering}
\subfloat[\label{fig:zero-a}]{\includegraphics[scale=0.3]{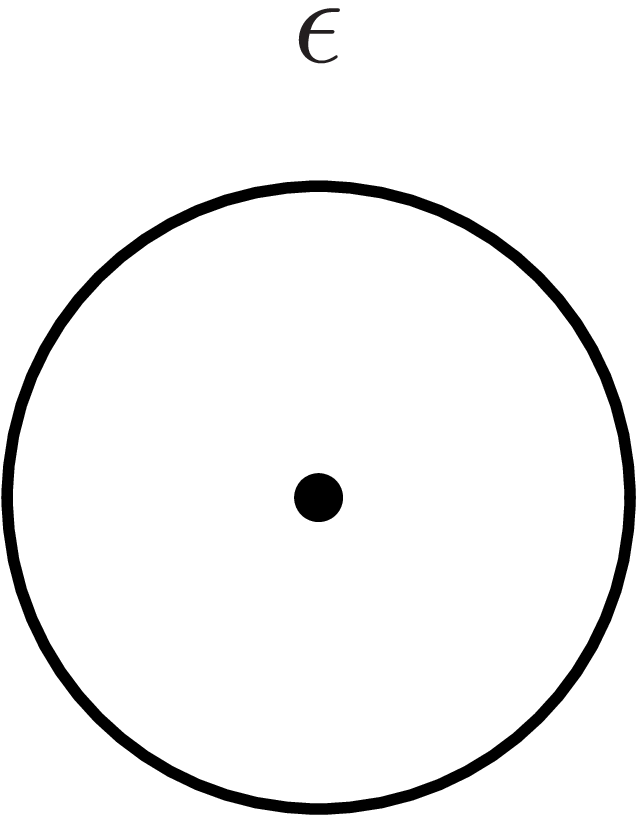}

} $\quad\quad\quad$\subfloat[\label{fig:zero-b}]{\includegraphics[scale=0.3]{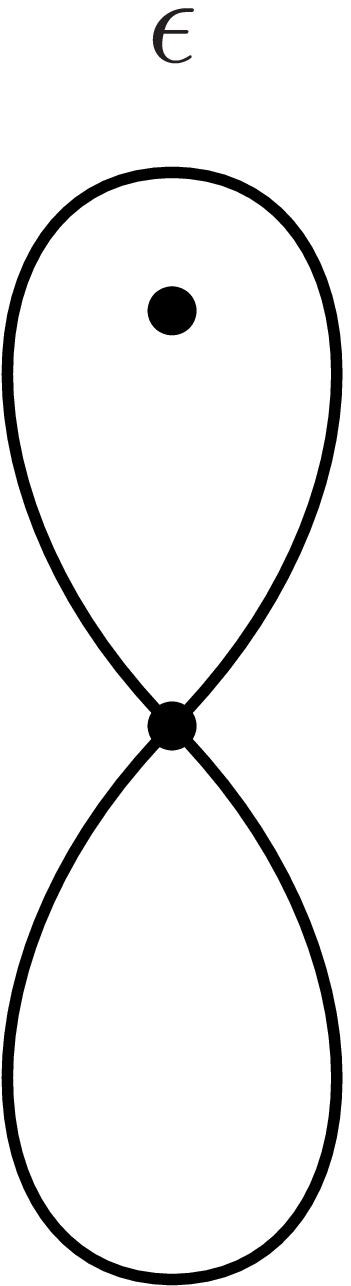}

}$\quad\quad\quad$\subfloat[\label{fig:zero-c}]{\includegraphics[scale=0.3]{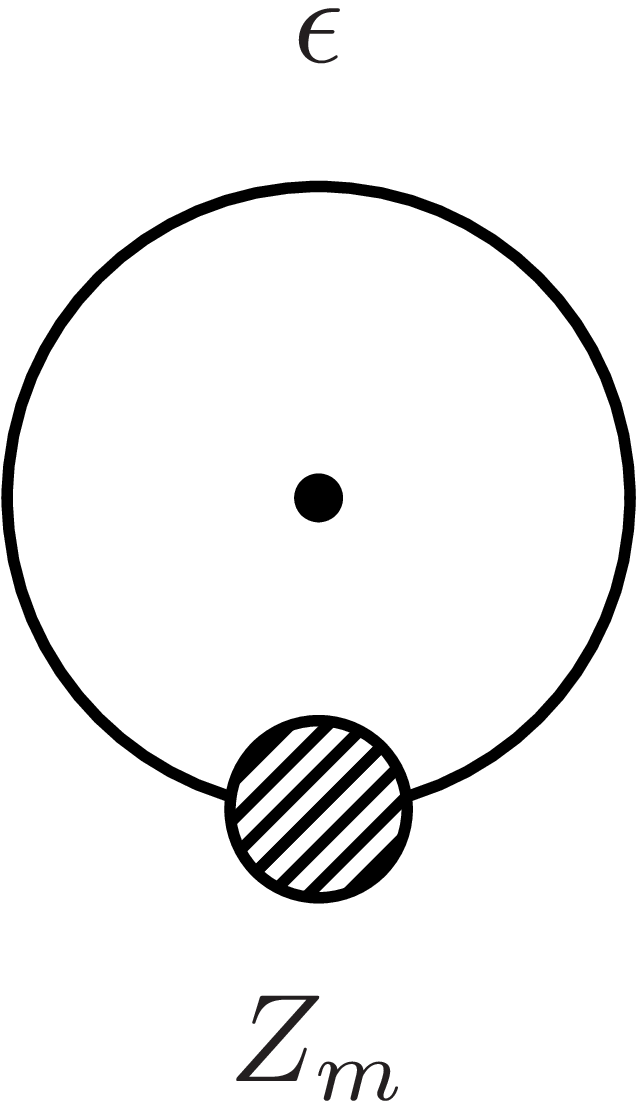}

}
\\
\subfloat[\label{fig:zero-d}]{\includegraphics[scale=0.3]{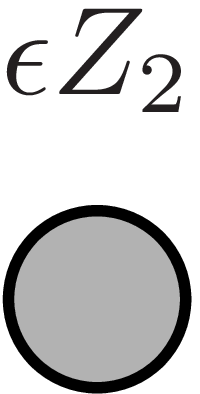}

}$\quad\quad$\subfloat[\label{fig:zero-e}]{\includegraphics[scale=0.3]{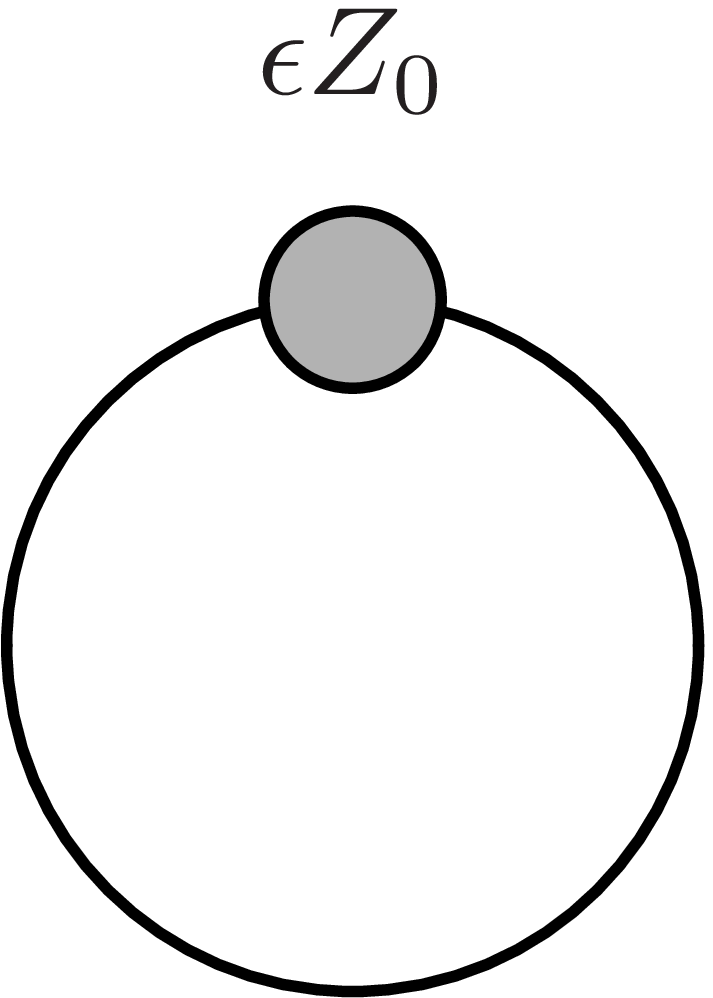}

}$\quad\quad$\subfloat[\label{fig:zero-f}]{\includegraphics[scale=0.3]{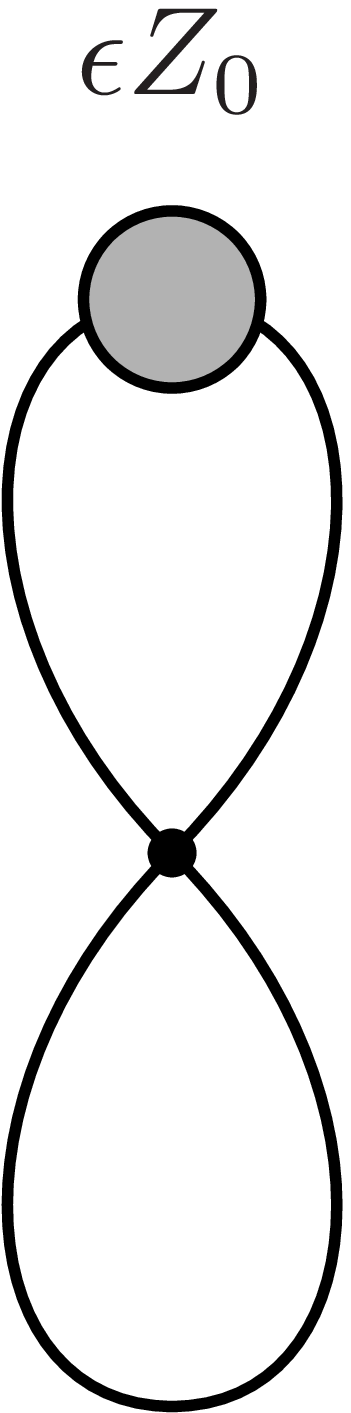}

}$\quad\quad$\subfloat[\label{fig:zero-g}]{\includegraphics[scale=0.3]{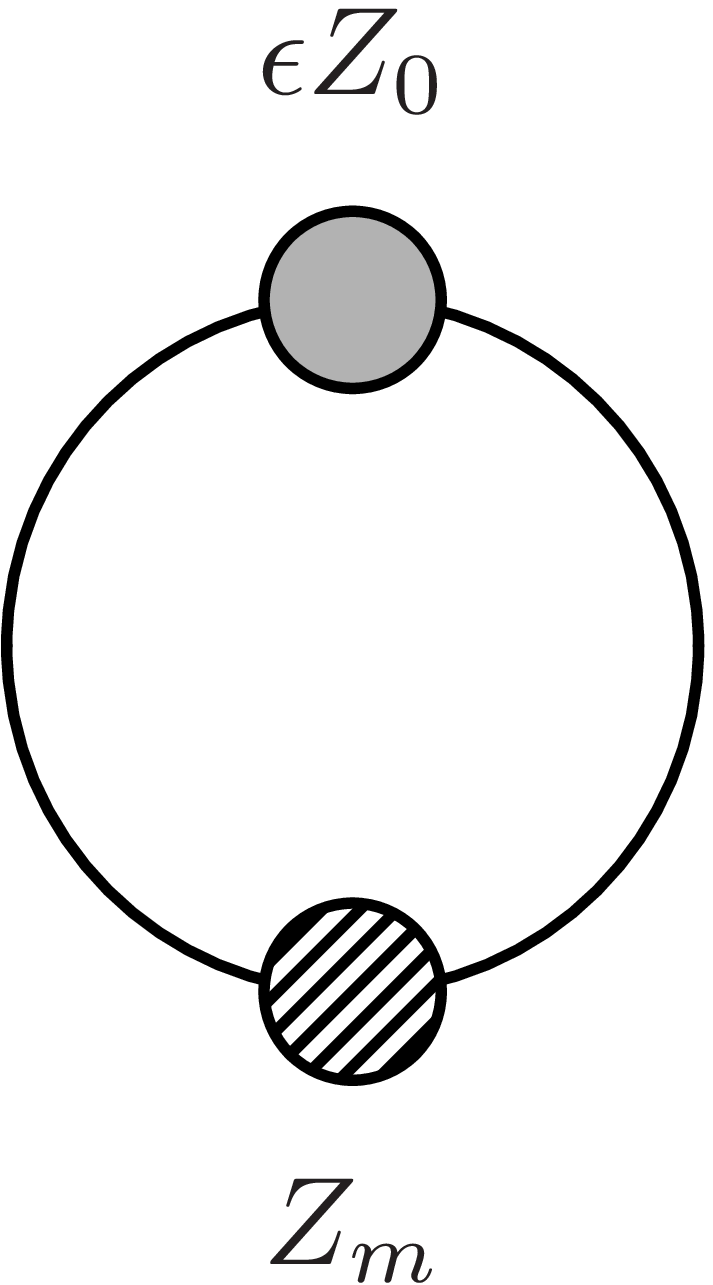}

}
\par\end{centering}
\captionsetup{justification=raggedright,singlelinecheck=false}

\caption{Zero point functions up to $O(\epsilon \lambda)$. Propagators circling around the dot are the $O(\epsilon)$ propagators. The gray blobs denote insertions of $Z_{0}$ or $Z_{2}$, while the shaded blobs denote insertions of $Z_{m}$.}
\label{fig:zero-point-function}

\end{figure}

Now it comes to the zero point function (the logarithm of the partition function) at $\mathcal{O}(\epsilon)$, which is exactly the entanglement entropy using the replica trick. Fig.(\ref{fig:zero-a}) denotes the contribution from the inverse determinant of the free theory. It can be computed by taking the $m^2$ derivative then integrate  $m^2$ back:
\begin{eqnarray}
\label{14}
 S_E^{(\ref{fig:zero-a})}
 =  \frac{1}{12}A_{\perp}\int_{m^{2}}^{\infty}d\mu^{2}\int\frac{d^{2}p_{\perp}}{(2\pi)^{2}}\frac{1}{p_{\perp}^{2}+\mu^{2}} +A_{\perp}S_{0}
\end{eqnarray}
$S_0$ is the contribution when $m \to \infty$. So $S_0$ is $m$ independent. 

We use the renormlization condition that the tree level mass is the physical mass already, such that the loop corrections to the mass are all canceled by counterterms.  Therefore, the one loop correction in Fig.(\ref{fig:zero-b}) is exactly cancelled by the insertion of the mass counterterm $Z_m$ in Fig.(\ref{fig:zero-c}):
\begin{eqnarray}
 S_E^{(\ref{fig:zero-b})+(\ref{fig:zero-c})} = 0 .
\end{eqnarray}
The $Z_2$ and $Z_0$ terms yield
\begin{eqnarray}
 S_E^{(\ref{fig:zero-d})}
 &=&  -A_{\perp} Z_2 ,  \\
 S_E^{(\ref{fig:zero-e})}
 &=&  -A_{\perp}Z_0\int\frac{d^{4}p}{(2\pi)^{4}}\frac{1}{p^{2}+m^{2}} .
 \label{16}
 \label{17}
\end{eqnarray}
And the one loop correction to (\ref{fig:zero-e}) vanishes
\begin{eqnarray}
 S_E^{(\ref{fig:zero-f})+(\ref{fig:zero-g})} =  0 .
\end{eqnarray}
If we take the mass derivative to Eqs.(\ref{14}) and (\ref{17}), the same combination as Eq.(\ref{2pt}) arises and yields
\begin{eqnarray}
S_E^{(\ref{fig:zero-a})+(\ref{fig:zero-e})}=-A_{\perp}\left(\frac{m^2}{48 \pi}+C\right) ,
\end{eqnarray}
where $C$ is $m$ independent. $C$ could diverge like $\Lambda^2$, with $\Lambda$ the ultraviolet momentum cut-off.

Putting everything together, we have 
\begin{eqnarray}
\label{20}
\frac{S_E}{A_{\perp}} = -\frac{m^2}{48 \pi} -\Bar{Z}_2+\mathcal{O}(\lambda^{2}),
\end{eqnarray}
where $\Bar{Z}_2=Z_2+C$.  The divergence in $C$ can be absorbed by $Z_2$ such that $\Bar{Z}_2$ is finite. Therefore, we have demonstrated that,  up to $\mathcal{O}(\lambda)$, with couplings $Z_0$ and $Z_2$ living in the two dimensional brane formed by the interface, $S_E$ could be properly renormalized to be a finite quantity.  

Here we have only discussed the $S_E$ of the ground state of a scalar field theory with an infinitely large flat interface to $\mathcal{O}(\lambda)$. But the conclusion that entanglement entropy is renormalizable could be general. One can generalize our derivation to all orders in $\lambda$ using the standard techniques, and try to work on different interfaces and theories.  
However, one can already see in our simple example that while $Z_0$ can be determined by the renormalization of the two point function, fixing $Z_2$ (or $\bar{Z}_2$ of Eq.(\ref{20})) requires knowing its dependence on the deficit angel in a condensed matter system which lives in a flat space. This is conceptually illusive. We will have more discussions on this in the next section.

\section{The Correspondence in the Black Hole Case}

It is instructive to reproduce the computation showing the equivalence between the black hole thermal entropy and entanglement entropy which mathematically has a one-to-one correspondence to our case in the previous section. 

The Euclidean action of a quantum scalar field in classical gravity is
\begin{eqnarray}
\mathcal{S} = \int d^4 x \sqrt{g} \mathcal{L} ,
\end{eqnarray}
and
\begin{eqnarray}
\label{LL}
\mathcal{L} & = & \mathcal{L}_{\phi} -\frac{R}{16 \pi G}+\frac{\alpha}{4 \pi} \phi^2 R +\mathcal{O}(R^{2}) .
\end{eqnarray}
$\mathcal{L}_{\phi}$ is the first two lines of Eq.(\ref{eq:o(n)_sigma_Lagrangian}) written in the general covariant form. $R$ is the Ricci scalar. The dimension four $\mathcal{O}(R^{2})$ terms include $R^2$, $R_{\mu\nu}^2$ and $R_{\mu\nu\rho\sigma}^2$. 

For an infinitely massive black hole, the Hawking temperature is zero and the horizon is a flat infinite plan. The space outside the horizon can be described by the Rindler space which is locally flat except at the origin. Therefore when one computes the black hole entropy using Eq.(\ref{S_BH}), or the black hole entanglement entropy across the horizon using Eq.(\ref{SE2}), the spacetime geometry is $C_{\epsilon} \times R^2$ in both cases---the same as the $S_E$ computation in flat space in the previous section.
Furthermore, we have $R=4 \pi \epsilon \delta^{(2)}(x_{\parallel})$ which makes Eq.(\ref{LL}) have the same form as  Eq.(\ref{eq:o(n)_sigma_Lagrangian}). These lead to $S_{BH}=S_E$ for the black hole.

We can set $\alpha=Z_0$ to renormalize the two point function in Fig.(\ref{fig:Two-point-function}). As for the terms of $\mathcal{O}(R^{2})$, although they are dimension four, they do not contribute until $\mathcal{O}(\epsilon^2)$ so they do not contribute to $S_E$ or $S_{BH}$.\footnote{The $\mathcal{O}(R^{2})$ can still contribute at $\mathcal{O}(\epsilon)$ when the curvature at $\mathcal{O}(\epsilon^0)$ is not zero as considered in Ref. \cite{Calabrese2004}.}

Following the same procedure as in the previous section, we have 
\begin{eqnarray}
\frac{S_{BH}}{A_{\perp}} = \frac{1}{4G}-C-\frac{m^2}{48 \pi}+\mathcal{O}(\lambda^{2}) =\frac{1}{4G_R}+\mathcal{O}(\lambda^{2}),
\end{eqnarray}
where the $1/4G$ term is the Bekenstein-Hawking entropy which is $\mathcal{O}(1/\hbar)$, but
interestingly it can be derived with just classical gravity. The $-C-m^2/48 \pi$ contribution is from quantum corrections to the black hole entropy starting at  $\mathcal{O}(\hbar^0)$. 
The mass independent term $C$ could diverge like $\Lambda^2$. But its divergence is absorbed by $1/G$ and the combination on the right hand side is matched to the renormalized quantity $1/4G_R$. 
(Here $G_R$ is related to the zero point function of $\phi$ with one power of $R$ dependence. This quantity is proportional to $S_E$ in this problem.) \cite{Susskind:1993ws,Susskind:1994sm}.
Finally, we have
\begin{eqnarray}
S_{BH} = S_E= \frac{A_{\perp}}{4 G_R} .
\end{eqnarray}
Hence both $S_{BH}$ and $S_E$ are shown, up to $\mathcal{O}(\lambda)$, to be the same for an infinitely massive black hole. And the entropy per horizon area is set by the Planck scale.

Most of the discussion in this section can be found in \cite{Susskind:1993ws,Susskind:1994sm,Larsen:1995ax}, except the part that $\alpha$ can be fixed by the two point function renormalization in Fig.(\ref{fig:Two-point-function}). We find that each term in our condensed matter case has a counter part in
the gravitational theory. This suggests that our formulation is sensible from the point of view of general relativity. 
Although the deficit angle is easy to imagine in a gravitational theory, it requires the ``off-shell action'' to describe a black hole away from its Hawking temperature to generate the deficit angle \cite{Carlip:1993sa}. This suggests that in the condensed matter system, the determination of the interface term $Z_2$ will require properties from the ``off-shell action'' that 
is not included in its usual ``on-shell action'' which can possibly be determined by scattering in flat 3+1 dimensions.

\section{Conclusion and Discussion}

Our study suggests that entanglement entropy is renormalizable and is a physical quantity. We have demonstrated the renormalizability of the entanglement entropy of the $\lambda \phi^4$ at order $\lambda$ when the 3+1 dimensional theory is separated into two regions by an infinitely flat 2-dimensional interface. Using the replica trick, the computation of the entanglement entropy across the interface can be carried out by putting the theory on a curved spacetime background. We have shown by an explicit computation at $\mathcal{O}(\lambda)$
that once all the relevant operators are included in the action, the theory and hence the  entanglement entropy, can be renormalized. 

We also reviewed the computation of black hole entropy and entanglement entropy across the horizon for an infinitely massive black hole and found that our non-gravitational calculation has a one-to-one correspondence to the black hole case. This suggests that our formulation is sensible from the point of view of general relativity. 

To renormalize the $S_E$ in a non-gravitational system, one uses the coupling $Z_2$ which live in the interface to absorb the infinities from loop diagrams. If we wish to make a prediction to $S_E$, then $Z_2$ needs to be fixed by other observables. However, this coupling exists in the ``off-shell action.'' It is not clear what would be a good way to fix it. However, even without knowing $Z_2$, one can still make predictions to combinations of entanglement entropies, e.g. mutual entropy \cite{Liu:2012eea,Casini:2013rba,Liu:2013una}, whose $Z_2$ dependence cancels.

There are some obvious directions for future work. One could generalize the renormalizability proof to all orders in $\lambda$, or generalize it to other theories, for examples, the standard model of particle physics. In the latter case, there are actually no relevant couplings at $\mathcal{O}(\epsilon)$ other than what we have already written down in Eq.(\ref{eq:o(n)_sigma_Lagrangian}) and our $\mathcal{O}(\epsilon)$ propagator result in Eq.(\ref{fn}) can be easily generalized to propagators for other fields.

\appendix

\section{The derivation of Eq.(\ref{fn})}
The Laplacian operator in the $C_{\epsilon} \times R^2$ space is
\begin{align}
\Delta_n= & \partial_{\perp}^{2}+\partial_{r}^{2}+\frac{1}{r}\partial_{r}+\frac{1}{n^{2}r^{2}}\partial_{\theta}^{2},
\end{align}
where $\theta=[0,2 \pi)$ and $n=1-\epsilon$. Its eigenfunction
\begin{align}
\label{A2}
\phi_{n}(p,x)= & \sqrt{2\pi}J_{|\frac{l}{n}|}(p_{\parallel}r)e^{il\theta}e^{ip_{\perp}x_{\perp}}
\end{align}
satisfies
\begin{align}
(-\Delta_{n}+m^2)\phi_{n}(p,x)= (p^{2}+m^2)\phi_{n}(p,x).
\end{align}

Then the Green's function $G_n(x,x')$ can be constructed via
\begin{align}
\label{A4}
G_{n}(x,x')= & \int\frac{d^{4}p}{(2\pi)^{4}}\frac{d^{4}p'}{(2\pi)^{4}}\phi_n(p,x)\tilde{G}_{n}(p,p')\phi_n^{*}(p',x'),\nonumber \\
\tilde{G}_{n}(p,p')= & \frac{(2\pi)^{4}\delta(p,p')}{p^{2}+m^{2}},
\end{align}
which yields Eq.(\ref{Gn}) and the integral over $p_{\theta}$, the conjugate momentum of $\theta$, is understood as the sum over all integer $l$. 

In Eq.(\ref{11}), a general function of $x-x'$ can be written in a similar way as Eq.(\ref{A4}) but in flat ($n=1$) space: 
\begin{align}
F(x-x')= & \int\frac{d^{4}k}{(2\pi)^{4}} e^{-i k(x-x')}\tilde{F}(k)\nonumber \\
= & \int\frac{d^{4}k}{(2\pi)^{4}}\phi_1^{*}(k,x)\tilde{F}(k)\phi_1(k,x').
\end{align}
Then we have 
\begin{align}
&\int d^{4}xd^{4}x'F(x-x')G_{n}(x,x')\nonumber \\
&=  \int\frac{d^{4}p}{(2\pi)^{4}}\frac{d^{4}k}{(2\pi)^{4}}\frac{F(k)}{p^{2}+m^{2}}\int d^{4}xd^{4}x'\nonumber \\
 & \ \ \ \times\phi_{n}(p,x)\phi_{n}^{*}(p,x')\phi_{1}^{*}(k,x)\phi_{1}(k,x') .
 \end{align}
We only need the $\mathcal{O}(\epsilon)$ contribution of this integral, which is
\begin{align}
\label{A7}
I&=\int d^{4}xd^{4}x'F(x-x')f(x,x')\nonumber \\
 &= -\int\frac{d^{4}pd^{4}x}{(2\pi)^{4}}\frac{F(p)}{p^{2}+m^{2}}\partial_{n \to 1}|\phi_{n}(p,x)|^2 ,
\end{align}
where we have used
\begin{align}
\int d^{4}x\phi^{*}(p,x)\phi(p',x)= & (2\pi)^{4}\delta^4(p-p').
\end{align}

Using Eq.(\ref{A2}), the $x$ integral in Eq.(\ref{A7}) is
\begin{align}
\label{A9}
&  \int d^{4}x \partial_{n \to 1}|\phi_{n}(p,x)|^2  \nonumber \\
& =A_{\perp} (2 \pi)^2 \partial_{n \to 1} \int dr r J^2_{|\frac{l}{n}|}(p_{\parallel}r) \nonumber \\
& =A_{\perp} (2 \pi)^2 \partial_{n \to 1} \frac{\delta(0)}{p_{\parallel}} .
\end{align}
The last expression vanishes unless $p_{\parallel}=0$. Since $A_{\perp}$ and $p_{\parallel}$ are the only scales in this expression, 
dimensional analysis suggests that it is proportional to $A_{\perp} \delta(p_{\parallel}^2)$. The proportional constant, $|l|$, can be fixed by considering the integral
\begin{align}
& \int\frac{pdp}{p^{2}+m^{2}}\partial_{n \to 1}\left[\int rdrJ_{|\frac{l}{n}|}^{2}(pr)\right]\nonumber \\
= & \partial_{n \to 1}\int rdrI_{|\frac{l}{n}|}(mr)K_{|\frac{l}{n}|}(mr)\nonumber \\
= & \frac{1}{m^{2}}\partial_{n \to 1}\left[-\frac{|l|}{2n}+\text{const.}\right]=\frac{|l|}{2m^2}.
\end{align}

Now we are ready rewrite Eq.(\ref{A7}) as 
\begin{align}
I&=-\int\frac{d^{4}p}{(2\pi)^{4}}\frac{F(p)}{p^{2}+m^{2}} |l| A_{\perp} \delta(p_{\parallel}^2) \nonumber \\
 &=-\int\frac{d^{2}p_{\perp}d p_{\parallel} p_{\parallel}}{(2\pi)^{4}}\sum_{-\infty}^{\infty}\frac{F(p)}{p^{2}+m^{2}} |l| A_{\perp} \delta(p_{\parallel}^2) \nonumber \\
 &=\frac{1}{12}\int \frac{d^{2}p_{\perp}}{(2\pi)^{4}}\frac{F(p_{\perp},p_{\parallel}=0)}{p_{\perp}^{2}+m^{2}} ,
 \end{align}
 where we have used the Riemann zeta function $\zeta(s)=\sum_{n=1}^{\infty} n^{-s}$ and $\zeta(-1)=-1/12$ via analytic continuation. The final result is summarized in Eq.(\ref{fn}).

\
\
\

\begin{acknowledgments}
The authors would like to thank Tatsuma Nishioka, Naoki Yamamoto,
Michael Endres, Xing Huang, Feng-Li Lin, Chen-Te Ma, Masahiro Nozaki,
Jackson Wu and Yun-Long Zhang for helpful discussions. 
JWC is partly supported by the Ministry of Science and Technology, Taiwan, under Grant Nos. 105-2112-M-002-017-MY3, 104-2923-M-002-003-MY3, the Kenda Foundation, and DFG and NSFC (CRC 110).
JYP is supported by DFG through funds provided to the Sino-German CRC 110 ``Symmetries and the Emergence of Structure in QCD''. 
\end{acknowledgments}


\end{document}